\documentclass[useAMS,usenatbib,aas_macros]{mn2e}

\usepackage{epsfig}
\usepackage{lscape,graphicx}
\usepackage{amssymb}
\usepackage{natbib}
\usepackage{times}
\usepackage{aas_macros}

\newcommand{\ifm}[1]{\relax\ifmmode#1\else$\mathsurround=0pt#1$\fi}
\newcommand{\kms}{\ifmmode\,{\rm km}\,{\rm s}^{-1}\else km$\,$s$^{-1}$\fi}
\newcommand{\Mpc}{\,{\rm Mpc}}

\newcommand{\hmsun}{\,\ifm{h^{-1}}{M_{\odot}}}
\def\omm{\Omega_{\rm m}}
\def\oml{\Omega_{\Lambda}}

\newcommand{\ltsima}{$\; \buildrel < \over \sim \;$}
\newcommand{\lsim}{\lower.5ex\hbox{\ltsima}}
\newcommand{\gtsima}{$\; \buildrel > \over \sim \;$}
\newcommand{\gsim}{\lower.5ex\hbox{\gtsima}}

\newcommand{\dd}{{\rm d}}

\newcommand{\equ}[1]{eq.~(\ref{eq:#1})}
\newcommand{\equs}[1]{eqs.~(\ref{eq:#1})}

\newcommand{\equnp}[1]{eq.~\ref{eq:#1}}
\newcommand{\se}[1]{\S\ref{sec:#1}}
\newcommand{\fig}[1]{Fig.~\ref{fig:#1}}

\newcommand{\Fig}[1]{Figure~\ref{fig:#1}}
\newcommand{\Figs}[1]{Figures~\ref{fig:#1}}

\newcommand{\dS}{\Delta S}
\newcommand{\dW}{\Delta \omega}

\newcommand{\be}{\begin{equation}}
\newcommand{\ee}{\end{equation}}
\newcommand{\bea}{\begin{eqnarray}}
\newcommand{\eea}{\end{eqnarray}}

\def\ltsima{$\; \buildrel < \over \sim \;$}
\def\lsim{\lower.7ex\hbox{\ltsima}}

\def\lan{\langle}
\def\ran{\rangle}

\def\mall{M_{\rm all}}
\def\mleft{M_{\rm left}}
\def\sleft{S_{\rm left}}

\begin{document}

\title[Constructing Merger Trees]
      {Constructing Merger Trees that Mimic $N$-Body Simulations}

\author[E.~Neistein \& A.~Dekel]
       {Eyal Neistein
        and Avishai Dekel\\
      Racah Institute of Physics, The Hebrew University,
          Jerusalem, Israel\\
     e-mails: eyal$\underline{\;\;}$n@phys.huji.ac.il;
              dekel@phys.huji.ac.il }


\date{}
\pagerange{\pageref{firstpage}--\pageref{lastpage}} \pubyear{2007}
\maketitle

\label{firstpage}


\begin{abstract}
We present a simple and efficient empirical algorithm for constructing dark-matter halo merger trees that reproduce the distribution of trees in the Millennium cosmological $N$-body simulation. The generated trees are significantly better than EPS trees. The algorithm is Markovian,  and  it  therefore  fails  to reproduce the non-Markov features of trees across short time steps, except for an accurate fit
to the evolution of the average main progenitor. However, it properly recovers   the  full  main  progenitor  distribution  and  the  joint
distributions  of  all  the  progenitors over long-enough time steps,
$\Delta  \omega \simeq \Delta z>0.5$, where $\omega \simeq 1.69/D(t)$
is  the  self-similar  time  variable and $D(t)$ refers to the linear
growth  of  density  fluctuations.  We  find that the main progenitor
distribution is log-normal in the variable $\sigma^2(M)$, the variance of linear density fluctuations in a sphere encompassing mass $M$. The
secondary  progenitors  are  successfully  drawn  one by one from the
remaining mass using a similar distribution function. These empirical
findings  may  be  clues  to  the  underlying  physics of merger-tree
statistics.  As  a  byproduct,  we  provide useful, accurate analytic
time-invariant approximations for the main progenitor accretion       history and for halo merger rates.

\end{abstract}




\section{Introduction}
\label{sec:intro}

Dark-matter (DM) haloes are the building blocks of non-linear structure in the universe.  They are the spheroidal gravitating systems in virial equilibrium within which galaxies form and live. The spherical collapse model in a cosmological background implies that the outer, virial radius of a halo can be defined by the radius
encompassing a mean overdensity of $\Delta \sim 200$ compared to the universal mean. The haloes are assumed to assemble hierarchically bottom-up, starting from Gaussian random initial density fluctuations that grow by gravitational instability and eventually detach from the expanding background, collapse and virilize. Most of the growth of a halo can be viewed as a sequence of mergers of haloes above an arbitrary minimum mass, termed ``progenitors", with the rest of the assembled mass considered ``smooth accretion".  The merger trees, describing the whole merger histories of DM haloes, serve as the backbone of galaxy formation.

The statistics of the halo distribution can be approximated by the
Press-Schechter formalism \citep{Press74}. Its extension \citep[EPS,
][]{Bond91,Lacey93} provides an approximate description of the
statistics of merger trees as a stochastic process in which the
probability for the set of progenitors is given. Both Press-Schechter \& EPS are based on the initial fluctuation power spectrum combined with the analytic model of cosmological spherical collapse. The EPS formalism is widely used in studies of structure formation \citep[e.g.][]{Lacey94, Mo96, Hernquist03}, especially through algorithms for the construction of random realizations of merger-trees \citep{Kauffmann93,Sheth99,Somerville99a,Cole00}. These algorithms enable detailed ``semi-analytic" simulations of galaxy formation models, as they are fast and allow a broad range of halo masses.

While the EPS trees are useful for semi-quantitative studies, their
accuracy may be insufficient for detailed comparisons with
observations. When compared to merger trees extracted from $N$-body
simulations, the EPS trees show non-negligible deviations, e.g., in
the number of progenitors \citep{Sheth02}, the growth history of the
main progenitor \citep{Wechsler02} and the mass contained within all
the progenitors \citep{Neistein06}. In addition, the EPS theory does
not uniquely define the full joint distribution of progenitor masses
and the associated merger rates \citep[e.g.][]{Somerville00}.
Therefore, different EPS-based algorithms may lead to trees with
different statistical characteristics and predict different merger
rates.

In terms of accuracy, $N$-body simulations should generate ``true"
merger trees. With the availability of large-volume simulations such
as the Millennium Run \citep{Springel05}, cosmic variance is no longer an issue. As a result, haloes in the mass range that is relevant for galaxy formation are well sampled. The accuracy of the DM dynamics is limited only by the numerical resolution of particle mass and gravitational force. However, non-trivial difficulties are involved in the process of identifying haloes \citep{Davis85, Bullock01, Springel01} and in linking them to their earlier progenitors \citep[e.g.][]{Springel05, Harker06}. The resultant trees may depend on the algorithms adopted for these tasks, which could be quite arbitrary. The freedom in defining $N$-body merger trees partly reflects uncertainties in the adopted halo definition and its possible variation as a function of time or mass. Several authors define the virial radius based on a mean overdensity $\Delta(z)$ that varies in time following the spherical top-hat model, while others use the radius $R_{200}$ based on a fixed $\Delta=200$ \citep[e.g.][]{Cole96, Wechsler02,Cohn07}. In fact, the whole concept of a ``virial radius" is put to some doubt by the finding that the virial kinematics extends far beyond the conventional halo radii around haloes that are significantly smaller than the non-linear clustering scale $M_*$ \citep{Prada06}.

A merger tree is a {\it Markov\,} chain if for any halo of a given mass $M$ at time $t$, the probability distribution of progenitors at any other time is fully determined by $M$ and $t$. In particular, in a Markov tree, the history of each halo within the tree does not depend on its future properties. Markov trees are therefore easy to handle. Using self-similar time variable, the tree can be fully constructed using fixed probabilities of progenitor masses across small time steps. Since the history of a halo in a Markov tree is independent of its future, the halo properties do not depend on the large-scale environment.

EPS trees are Markovian if the haloes are defined by a convolution
with a top-hat window in Fourier space, but any other window
introduces correlations between the fluctuations on different scales,
which lead to non-Markov trees. This was first formulated by
\citet{Bond91}, and implemented, e.g., by \citet{Amosov04,Zentner07}.
Indeed, $N$-body trees are in general non-Markovian, making their
statistical description more complicated. This is evident from the
detection of environment dependence in halo histories that are
extracted from cosmological simulations \citep{Gao05,Harker06}. These
non-Markov features may arise from the correlations introduced by the smoothing of the initial density field, from the finite relaxation time associated with the non-linear assembly process, and from tidal effects including tidal stripping of haloes as they move near or inside other haloes \citep{Diemand07,Desjacques07,Hahn07}. We note that the deviations from a Markov behaviour may depend on the algorithm used to construct the merger trees.

Our goal here is to develop a simple and practical Markov algorithm for constructing merger trees, that will be easy to implement across
short time steps, and will provide a good fit to the statistics of
$N$-body merger trees once considered across large enough time steps.
We will find that this is a doable task once we identify the natural
variables of time and mass, which permit time-invariance and a robust functional shape for the distribution of progenitors in all halo masses. We aim to demonstrate that this algorithm provides a substantially better fit to the $N$-body trees than the EPS-based algorithms. A related analysis is provided independently by \citet{Parkinson07}, based on a different method and somewhat different merger trees.

The outline of this paper is as follows. In \se{millen_sim} we briefly describe the Millennium Run and the merger trees used. In \se{main_prog} we extract the main-progenitor history from the simulation, and show how we reproduce it with a Markov process. In \se{full_trees} we present the algorithm for constructing full merger trees and demonstrate that it is a significant improvement over EPS trees. In \se{merger_rates} we examine merger rates and mutual probabilities between progenitors. In \se{markov_limitation} we discuss the limitations of a Markov model. Finally, in \se{discuss}, we summarize our results and discuss them.


\section{The Millennium Simulation}
\label{sec:millen_sim}

Merger trees are obtained from the Millennium Run $N$-body simulation
\citep[][hereafter MR]{Springel05}, carried out by the Virgo Consortium. The cosmology is assumed to be $\Lambda$CDM, with the cosmological parameters $(\oml,\, \omm,\, \sigma_8,\, h) =(0.75,\, 0.25,\, 0.9,\, 0.73)$. The simulation follows the evolution of $2,160^3$ dark matter particles in a periodic box of a comoving side $500 h^{-1}$ Mpc from $z = 127$ to the present epoch. The particle mass is $8.6 \times 10^8\,\hmsun$, and the gravitational force has a comoving softening length of $5 h^{-1}$ kpc. The particle data were stored at 64 times, most of which are equally spaced in $\log(1+z)$ between $z = 20$ and 0. These output snapshots were then used for constructing merger trees.

We use the merger trees constructed from the MR using an FOF algorithm as described in \citet{Harker06}. This algorithm is suitable here because it focuses on distinct haloes that are not subhaloes of bigger haloes. FOF trees are especially appropriate for our purpose because such $N$-body trees were compared to EPS trees in the past \citep[e.g., ][]{Lacey94}. In practice, the total mass associated with a halo is first estimated using a linking length of $b=0.2$ compared to the mean near-neighbour distance. The mass estimate is then modified slightly in order to properly handle substructure, and some haloes are actually split when the automatic FOF linking seems unreasonable based on certain criteria \citep[see][for more details]{Harker06}. A non-standard procedure in the construction of merger trees is that the search for the descendant halo of a given halo is pursued over the subsequent five snapshots. Such subtle details of the halo finder and the tree-construction algorithm may have non-negligible effects on the statistics of the merger trees.

For most purposes we focus on haloes that are identified at the present epoch, $z=0$. We divide these haloes according to their final mass $M_0$ at $z=0$ (the tree ``trunk") into four representative bins, as listed in table \ref{tab:halo_masses}. The bins become broader at larger masses to ensure a sufficient number of haloes in each bin for good statistics.

\begin{table}
\caption{The merger trees are divided into 4 bins according to the mass $M_0$ of the final halo at $z=0$. Each row of the table refers to a different bin, defined by $M_{\rm low} \leq M_0 \leq M_{\rm high}$, and consisting of $N$ trees. All masses are in units of $\hmsun$.}
\begin{center}
\begin{tabular}{lccc}
\hline Average mass& $M_{\rm low}$ & $M_{\rm high}$ & $N$ \\
\hline
$10^{11}$ & $10^{11}$ & $1.05\times10^{11}$ & $3\times10^5$ \\
$1.4\times10^{12}$ & $10^{12}$ & $2\times10^{12}$ & $2\times10^5$ \\
$2\times10^{13}$ & $10^{13}$ & $5\times10^{13}$ & $4\times10^4$ \\
$2.1\times10^{14}$ & $10^{14}$ & $10^{15}$ & $3\times10^3$ \\
\hline
\end{tabular}
\end{center}
\label{tab:halo_masses}
\end{table}
%


\section{Main Progenitor History}
\label{sec:main_prog}

The history of the ``main progenitor" (hereafter MP) is constructed by following back in time the most massive progenitor in each merger event. The mass growth of the MP is interpreted for certain purposes
as the mass growth history of the final halo, e.g., when identifying
a characteristic assembly time for the halo. This has been useful in quantifying important aspects of the merger histories of haloes \citep{Lacey93, Wechsler02, vdBosch02b, Li07}, and helped in the understanding of certain issues concerning galaxy formation \citep[e.g.][]{vdBosch02a, Birnboim07}. We denote by $P_1(M_1|M_0,z,z_0)$ the conditional probability to have at $z$ a MP of mass $M_1$, given that it has merged by $z_0$ into a halo of mass $M_0$. We will investigate below to what extent $P_1$ can be fitted by a unique log-normal distribution function, for all masses and at all times. In particular, we will study the requirements from the length of the time step for this to be a good approximation. This will allow us to construct the full statistics of the merger history using a Markov chain model.


\subsection{Natural Variables}
\label{sec:natural_var}

The first key step is to identify a natural time-variable $\tau$ under which the trees are time-invariant. In particular, we wish $P_1$ to
depend only on $\Delta \tau = \tau(z)-\tau(z_0)$ and be independent of $z_0$. The natural time variable emerging from the EPS theory is
$\omega \equiv \delta_{c}(z)/D(z)$, where $\delta_c(z)\simeq 1.69$
with a weak dependence on $z$ and $D(z)$ is the cosmological linear growth rate (see appendix \ref{sec:app_w_s}). Any time dependence in EPS trees enters only through $\dW=\omega(z)-\omega(z_0)$. Indeed, previous analytical derivations of MP ``formation time" \citep{Lacey93} and the full average mass history \citep{Neistein06}, based on EPS, used $\omega$ as the time variable. Alternatively, one could try $z$ as the time variable. This led \citet[][based on EPS trees]{vdBosch02b} and \citet{Wechsler02} to formulae for the average MP history in good agreement with earlier $N$-body simulations for haloes that are identified at $z_0=0$. \citet{Wechsler02} also proposed that $\Delta z$ allows a good time-invariant generalization to other $z_0$ measurement times\footnote{One should correct a typo in eq.~5 of \citet{Wechsler02}, where $a_c$ should be replaced by $a_c/a_0$ for $a_c$ to be the formation time as defined there, and independent of $a_0$. Note also that they defined haloes based on $\Delta(z)$ within a sphere, while the MR haloes are based on FOF with a constant $b=0.2$.}.
We next test to what extent these time variables lead to time
invariance of the MP distribution $P_1$ in the Millennium Run.

\begin{figure}
\centerline{ \hbox{ \epsfig{file=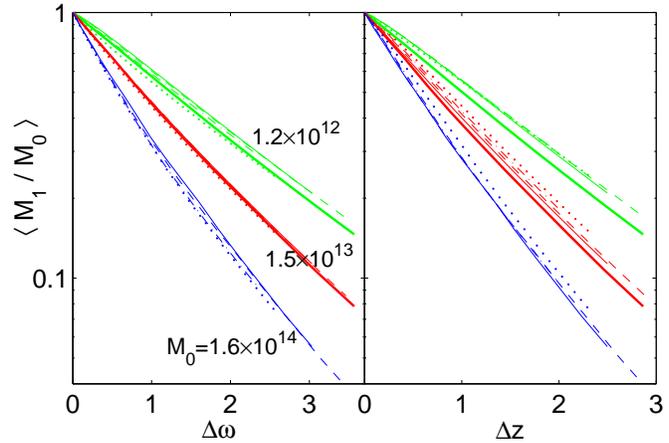,width=9cm} }}
\caption{Time invariance of the average MP history for two different
time variables, $\Delta \omega$ and $\Delta z$. The three bundles of
curves refer to three different halo masses: $M_0=1.2\times10^{12},
1.5\times10^{13}$, $1.6\times10^{14}\;\hmsun$ (green, red, blue from
top to bottom). The curves in each bundle refer to
$z_0=0,0.4,0.7,1.1,2.4$ (dotted, dashed-dotted, dashed, thin solid,
thick solid line types, except for the most massive haloes, where the
statistics is insufficient at $z_0=2.4$). In accordance with the EPS
spirit, the use of $\dW$ provides good time invariance, with
deviations of less than 10\% in $\lan M_1 \ran$, somewhat better than
the scatter for $\Delta z$. The scatter for the less massive haloes
may be affected by an environment effect, and for high mass haloes
sampling noise is dominant.}
  \label{fig:time_var}
\end{figure}

\Fig{time_var} shows the average MP history for haloes of mass $M_0$
at $z_0$, where $z_0$ is ranging from 0 to 2.4 for each given halo
mass. These histories are shown as a function of $\Delta \omega$ and
as a function of $\Delta z$. We see that $\dW$ provides good time
invariance, with a scatter of less than 10\% in the MP mass between
different $z_0$ values. We also see that the use of $\Delta z$ leads
to a reasonable time invariance, but with a somewhat larger scatter of $\lesssim20\%$, and with a stronger trend of increasing scatter at
earlier times. We report that we verified a similar time invariance
for other tree quantities, such as the number of progenitors and the
mutual probabilities of the two most massive progenitors. The above
has been verified for the $\Lambda$CDM cosmology used in the current
simulation.

It would be interesting to identify the source of residual scatter in the average MP mass when $z_0$ is varied and the time variable
$\Delta \omega$ is used. This scatter could have potentially been an artifact of the redshift dependence of the time steps used in the construction of the merger trees. For example, the MP may be the most massive progenitor or not depending on the length of the time step. In order to test this, we used the $z_0=0$ haloes to compare $P_1$ at $\dW\sim0.4$ as produced using different time steps corresponding to $\dW$ ranging from 0.016 to 0.4. We find the resultant scatter to be negligible. A more relevant source of scatter is the environment dependence of halo formation time \citep{Gao05,Harker06}, detected as a weak correlation between the redshift at which $\lan M_1 \ran = 0.5M_0$ and the environment density for a given $M_0$. This is especially true for haloes of masses $M_0 \ll M_*$, where $M_*(z)$ is the Press-Schechter characteristic mass of nonlinear clustering. This may affect the curves in \fig{time_var} because the typical environment of haloes of a fixed mass is expected to vary with $z_0$. This is likely to be a significant source of scatter for the low masses, $M_0 \sim 10^{12}\,\hmsun$. Poisson noise is important only in the massive halo bin, where the number of haloes decrease from $\sim 3000$ at $z_0=0$ to $\sim 300$ at $z_0=1.1$ (note that the mass bins used here are not the ones described in table \ref{tab:halo_masses}). In order to test this noise we made 1000 runs of merger trees using our algorithm as described below. Each run contained 300 trees, for which the average mass was computed, similarly to the MR sample. The standard deviation between all theses averages gives a non-negligible error of $\sim 4\%$ at $\dW=2$. This error is comparable to the scatter we see between different $z_0$.

\begin{figure}
 \centerline{ \hbox{ \epsfig{file=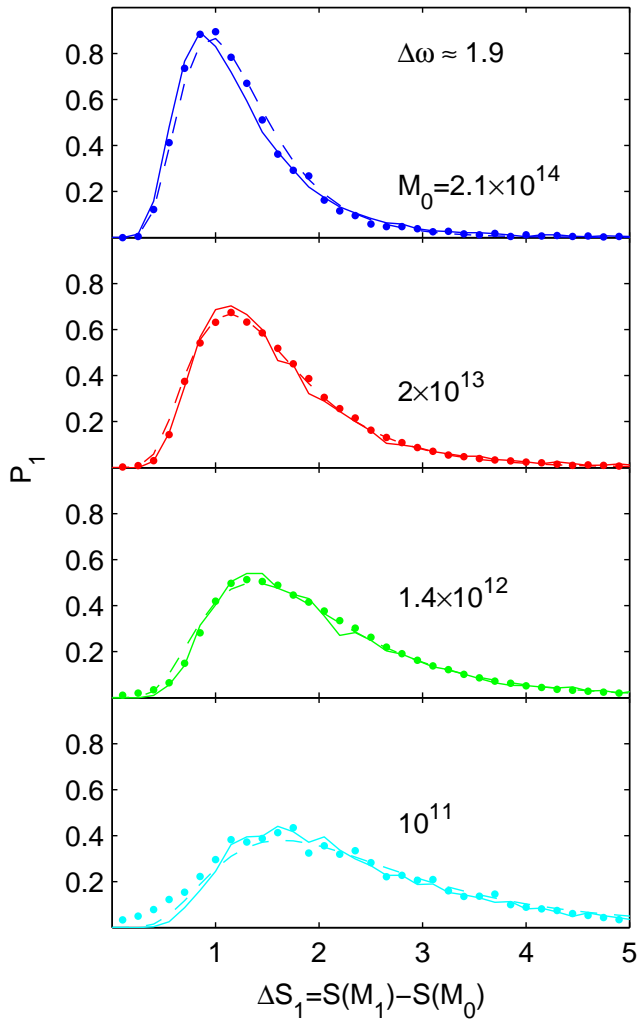,width=10cm} } }
\caption{The probability distribution of the MP mass,
$P_1(\dS_1|S_0,\dW)$. Halo masses, $M_0$ at $z=0$, correspond to the
mass bins listed in table \ref{tab:halo_masses}. The mass difference,
$\dS_1=S(M_1)-S(M_0)$, refers to the main progenitor back at the time corresponding to $\dW=1.9$. Shown for each halo mass is the distribution deduced from the Millennium Run (filled circles), along with the global fit by the log-normal distribution of
\equs{logn_fit1}-(\ref{eq:logn_fit3}) (dashed curve). Shown in comparison (solid curve) is the distribution from $10^4$ random realizations of merger trees generated by the algorithm described in \se{markov_model}, using the same distribution of $M_0$ as in the MR.}
  \label{fig:main_prog_model}
\end{figure}

\begin{figure}
 \centerline{ \hbox{ \epsfig{file=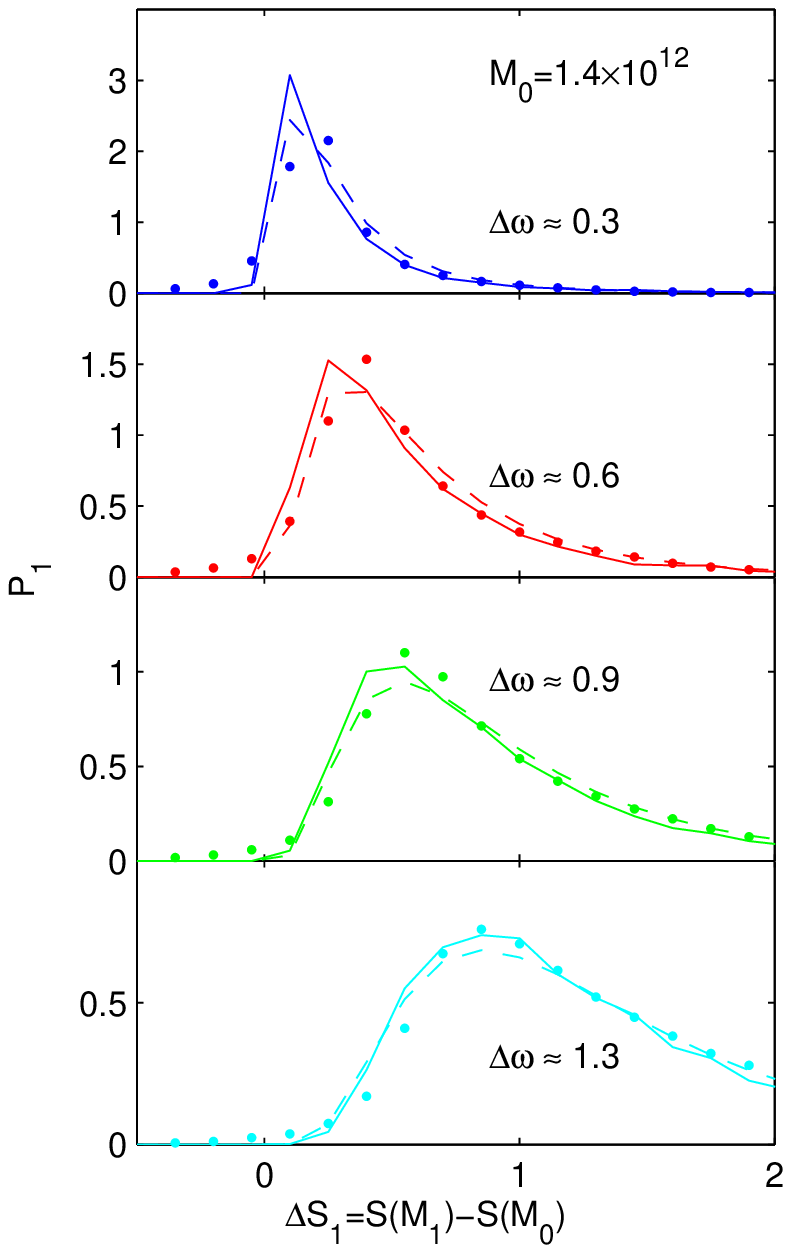,width=10cm} } }
\caption{Similar to \fig{main_prog_model}, but for the fixed mass bin
$1.4\times10^{12}\,\hmsun$ and different time steps $\dW$ as indicated. Note the different scaling of $P_1$ in the different panels.}
  \label{fig:main_prog_with_t}
\end{figure}

Next we wish to identify a mass variable that would make $P_1$ fit by
a simple functional form, the same for all masses. An immediate choice could have been $M_1/M_0$, but we could not find a simple functional form involving this variable that would provide a good robust fit to the simulation. Instead, we test $\dS_1=S(M_1)-S(M_0)$, where $S(M)=\sigma^2(M)$ is the variance of the initial density fluctuation
field, linearly extrapolated to $z=0$, and smoothed using a window
function that corresponds to a mass $M$. This is the natural mass
variable used in EPS \citep{Lacey93}. Note that the natural time
variable and this mass variable are related via $\omega(z)=\sigma[M_*(z)]$, which serves as the definition of the
Press-Schechter mass $M_*$. We describe how we compute $S(M)$ in
appendix \ref{sec:app_w_s}.

\Figs{main_prog_model} and \ref{fig:main_prog_with_t} focus
on the MP distribution $P_1(\dS_1|S_0,\dW)$ at several times $\dW$
and for different fixed halo masses $M_0$ at $z_0=0$. The units of $P_1$ are $\dS^{-1}$, so its integral over $\dS$ equals
unity. We see that once the time-step is sufficiently long, $\dW>0.5$, the simulated distribution resembles a log-normal distribution in $\dS_1$,
\be
\label{eq:logn_fit1}
P_1(\dS_1|S_0,\dW) = \frac{1}{\sigma_p \dS_1
\sqrt{2\pi}} \exp\left[{ -\frac{(\ln\dS_1-\mu_p)^2}{2\sigma_p^2} }\right] \,.
\ee
The moments are expressed as functions of $M_0$ and $\dW$,
\begin{eqnarray}
\label{eq:logn_fit2}
\sigma_p \!\!\!\!&=&\!\!\!\!
\left( a_1\log_{10}M_0 + a_2 \right) \log_{10} \dW +a_3\log_{10}M_0 + a_4 \;,\\
\nonumber
\mu_p \!\!\!\!&=&\!\!\!\!
\left( b_1\log_{10}M_0 + b_2 \right) \log_{10} \dW + b_3\log_{10}M_0 + b_4 \;,
\end{eqnarray}
and the best-fit parameters are determined once, globally for all halo masses and times as listed in table \ref{tab:halo_masses},
\begin{eqnarray}
\label{eq:logn_fit3}
(a_1,a_2,a_3,a_4)\!\!\!\!&=&\!\!\!\!(-4.5\times10^{-3},-0.34,-0.034,1.04) \;,\\
\nonumber
(b_1,b_2,b_3,b_4)\!\!\!\!&=&\!\!\!\!(0.072,1.56,-0.22 ,2.54) \;,
\end{eqnarray}
for $M_0$ is in units of $\hmsun$.
We discuss in appendix \ref{sec:app_mmain_fit} the quality of the global fit, and demonstrate that it improves with increasing time-steps, reaching at $\dW>1$ an accuracy of $\sim 20\%$ for the first four moments of the distribution. Deviations from the global fit are apparent in the figure mainly for the lowest-mass haloes, where the minimum mass resolution is not negligible. The fits can obviously be improved further once the parameters are determined separately for each halo mass.


\subsection{A Markov-Chain Model}
\label{sec:markov_model}

We wish to generate random MP histories in a simple way through a sequence of equal, small time-steps, $\dW_0$, that sum up to the desired long time-step $\dW$. In each time-step $i$, we draw a random
mass-step $\dS_{1,i}$ from a fixed kernel probability function $K_1(\dS_{1,i}|S)$. The probability $P_1(\dS_1|S_0,\dW)$ is obtained
by summing up the small mass-steps $\dS_{1,i}$ over all the time-steps. The Markov chain is thus:
\begin{eqnarray}
\lefteqn{ \dS_1(S_0) = \dS_{1,1}(S_0) + \dS_{1,2}(S_0+\dS_{1,1}) + } \\
\nonumber & & \ldots + \dS_{1,n}\left( S_0 + \sum_{i=1}^{n-1}\dS_{1,i}
\right) \;.
\end{eqnarray}
For example, in the case of two steps,
\begin{eqnarray}
\lefteqn{ P_1(\dS_1|S_0,2\dW_0) = } \\ \nonumber & & \int_0^{\dS_1}
K_1(\dS|S_0)\; K_1(\dS_1-\dS|S_0+\dS) \dd \dS \;.
\end{eqnarray}

A simple solution might have been to use $K_1=P_1(\dS_1|S_0,\dW_0)$ as extracted directly from the Millennium Run. However, this procedure fails because the MR trees are not Markovian for small time-steps, namely $P_1$ also depends on the $\dS$ of previous time-steps. Consequently, the $P_1$ from the MR is usable only for large time-steps, $\dW_0\gtrsim0.5$. Such large time-steps are not good enough for certain applications which require a higher resolution in the merger history. In particular, long time steps involve multiple mergers, which have to be ordered in time for a proper evaluation of the merger rate (see \se{merger_rates}).

Our approach here is to assume a hidden Markov process that is
valid also for short time-steps. By applying its kernel $K_1$ over a
sequence of short time-steps, we wish to recover the MP distribution
in the MR at a big time-step. We should emphasize that $K_1$ is not
$P_1$, and is not obtained from the log-normal fit to $P_1$.
Nonetheless, we find that a suitable kernel is also provided by a
log-normal function,
\begin{eqnarray}
\label{eq:K(dS)}
\nonumber
K_1(\dS|S) \!\!\!\!&=&\!\!\!\! \frac{1}{\sigma_k \dS
\sqrt{2\pi}} \exp\left[{ -\frac{(\ln\dS-\mu_k)^2}{2\sigma_k^2} }\right] \,, \\
\sigma_k \!\!\!\!&=&\!\!\!\! 1.367 +0.012s + 0.234s^2 \,, \\
\nonumber
\mu_k \!\!\!\!&=&\!\!\!\! -3.682+0.76s - 0.36s^2 \,,
\end{eqnarray}
where $s\equiv \log_{10}(S)$. The best-fit parameters of $K_1$ were
derived using a Monte-Carlo search scheme, optimizing the fit to the
simulation data (table \ref{tab:halo_masses}) for $\dW>0.8$. Throughout this work, quite arbitrarily, we apply $K_1$ with a time-step $\dW_0=0.1$. We verified that any time-step in the range $0.01<\dW<0.2$ can yield a similar success. However, we failed to match the MR data with time steps as small as $\dW \sim 10^{-4}$, either because of a numerical effect or due to a more fundamental issue.

\begin{figure}
 \centerline{ \hbox{ \epsfig{file=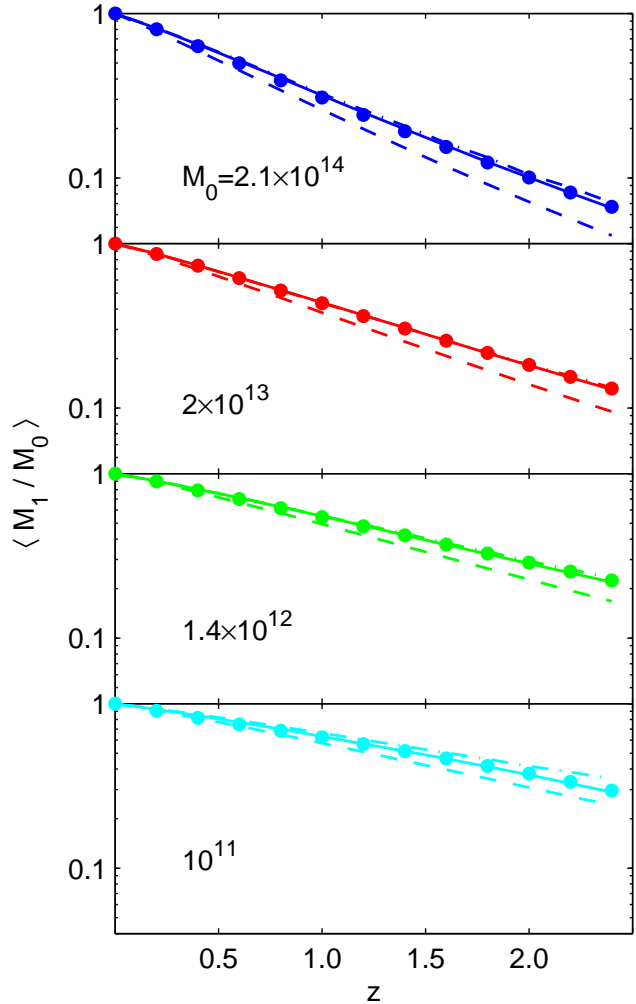,width=10cm} } }
\caption{Average mass of the MP $M_1$ at redshift $z$ for haloes with
mass $M_0$ at $z_0=0$ in the mass bins of table \ref{tab:halo_masses}. The results from the MR are marked by solid circles. The averages from merger-tree realizations as generated by our Markov model are the
solid curves. The analytic fit of \equ{average_mmain} gives rise to
the dot-dashed curves. The EPS predictions based on the analytic formula of \citet{Neistein06} are plotted as dashed lines. The Markov model and the analytic fit provide good fits to the data. The analytic fit is not as good for the $M_0=10^{11}\,\hmsun$ mass bin because
\equ{average_mmain} does not take into account the minimum halo mass
of the simulation.  The EPS fit is not as good.}
  \label{fig:main_prog_average}
\end{figure}

It should be emphasized that our model kernel $K_1$ guarantees that
the mass of the main progenitor is monotonically increasing with time
(namely $M$ is always decreasing with $\omega$), while this is not always true in the MR (see the small tail of $\dS<0$ in \fig{main_prog_with_t}). This may be important for semi-analytic models of galaxy formation, where the recipes for the baryonic processes become more complicated when halo mass loss occurs.

\Figs{main_prog_model} and \ref{fig:main_prog_with_t} compare the MP
distribution as generated from realizations of our Markov model to that deduced from the MR at several masses and times. For $\dW>1$ and for the mass range tested here, the model recovers the data at the level of $\sim 20\%$ in terms of the first four moments of the distribution (Appendix \ref{sec:app_mmain_fit}). For the highest mass bin, the accuracy of the fit at high $\dW$ is actually comparable to the simulation sampling noise. At short time-steps, $\dW<0.8$, the deviations of the model from the data tend to be larger.

While the model predictions of $P_1$ deviate from the $N$-body data at small time steps, the model manages to reproduce the \emph{average}
mass of the MP quite accurately even at small time steps. This is
demonstrated in \fig{main_prog_average}, which compares the average
mass of the MP by our Markov model with the data from the MR. The
fit is excellent for all masses and at all time-steps. The deviations
are below the $\sim 1\%$ level, much less than the scatter due to the deviations from time invariance when using $\dW$. Also shown in \fig{main_prog_average} are the predictions from the EPS model, as
computed by the analytic formula proposed by \citet{Neistein06}. Our
Markov model clearly performs much better than the EPS model.

The Markov model presented here allows us to construct very
efficiently many random realizations of the MP history. In particular, the transformation from $S$ to $M$, which is a demanding part of the
computation, needs to be performed only at a small number of output
times. Given that the log-normal distribution can be generated very
efficiently, we were able to produce MP histories at a rate of $\sim 10^4$ per second using a standard $\sim 1$GHz computer.


\subsection{Average Mass Accretion Histories}
\label{sec:average_mmain}

For practical purposes, it is useful to provide a simple fitting
formula that properly approximates the evolution of the average mass
of the main progenitor in the Millennium Run. We use a functional form similar to the one describing the MP in the EPS theory \citep{Neistein06}, in which the growth rate is given by
\be
\label{eq:average_mmain}
\frac{\dd M_{12} }{\dd \omega} = -\alpha
M_{12} ^{1+\beta}\;,
\ee
and the corresponding mass growth function is
\be M_{12}(\dW|M_0) = (M_{0,12}^{-\beta} +\alpha\beta\dW )^{-1/\beta}
\;, \label{eq:average_mmain2} \ee
where $M_{12}\equiv\lan M_1 \ran/10^{12}\hmsun$, with $\lan M_1 \ran$ the average mass of the MP, and where $M_{0,12}\equiv M_0/10^{12} \hmsun$. The best-fit parameters are $\alpha=0.59$ and $\beta=0.141$.

In order to express the growth rate in terms of time we write, $\dd
M_1/\dd t = \dot{\omega}\,\dd M_1/\dd \omega$. Recall that when approximating $\delta_c=const.$, $\dot\omega$ is given by $\dot \omega /\omega = - \dot D/D$. For a better accuracy, we offer here a simple explicit approximation for $\dot{\omega}$,
\begin{eqnarray}
\label{eq:dwdt} \dot{\omega} = -0.0470\left[ 1+z+0.1(1+z)^{-1.25}
\right]^{2.5} \; h_{73}\, {\rm Gyr}^{-1}\;,
\end{eqnarray}
where $h_{73}$ is the Hubble constant measured in units of 73
$\kms\,\Mpc^{-1}$. This approximation is valid in $\Lambda$CDM with
$\omm=0.25$ and $\oml=0.75$ to better than 0.5\%.

The fitting function of \equ{average_mmain2} is compared to the MR
data in \fig{main_prog_average}. The fit is good to better than 3\%
for the halo mass range studied and for $\dW<2.4$. Based on the time
invariance associated with $\dW$, as discussed in \se{natural_var},
one can straightforwardly extrapolate the fitting formula for the MP
history at higher redshifts. For example, the early history
($z\gtrsim2$) of the MP of a halo of $10^{14}\hmsun$ at $z_0=0$ is
similar to the recent history ($z \gtrsim 0$) of a $10^{13}\hmsun$
halo, once expressed in terms of $\dW$. The accuracy of the fitting
formula is limited by the accuracy of the time invariance associated
with $\dW$, \fig{time_var}. The apparent deviation of the simulation
data from the fitting formula for the halo mass of $M_0 \sim
10^{11}\,\hmsun$ and below stems from the minimum halo mass imposed in the simulations, $M_{\rm min} = 1.72\times 10^{10}\,\hmsun$.


\section{Constructing Full Trees}
\label{sec:full_trees}

\begin{figure*}
\centerline{\psfig{file=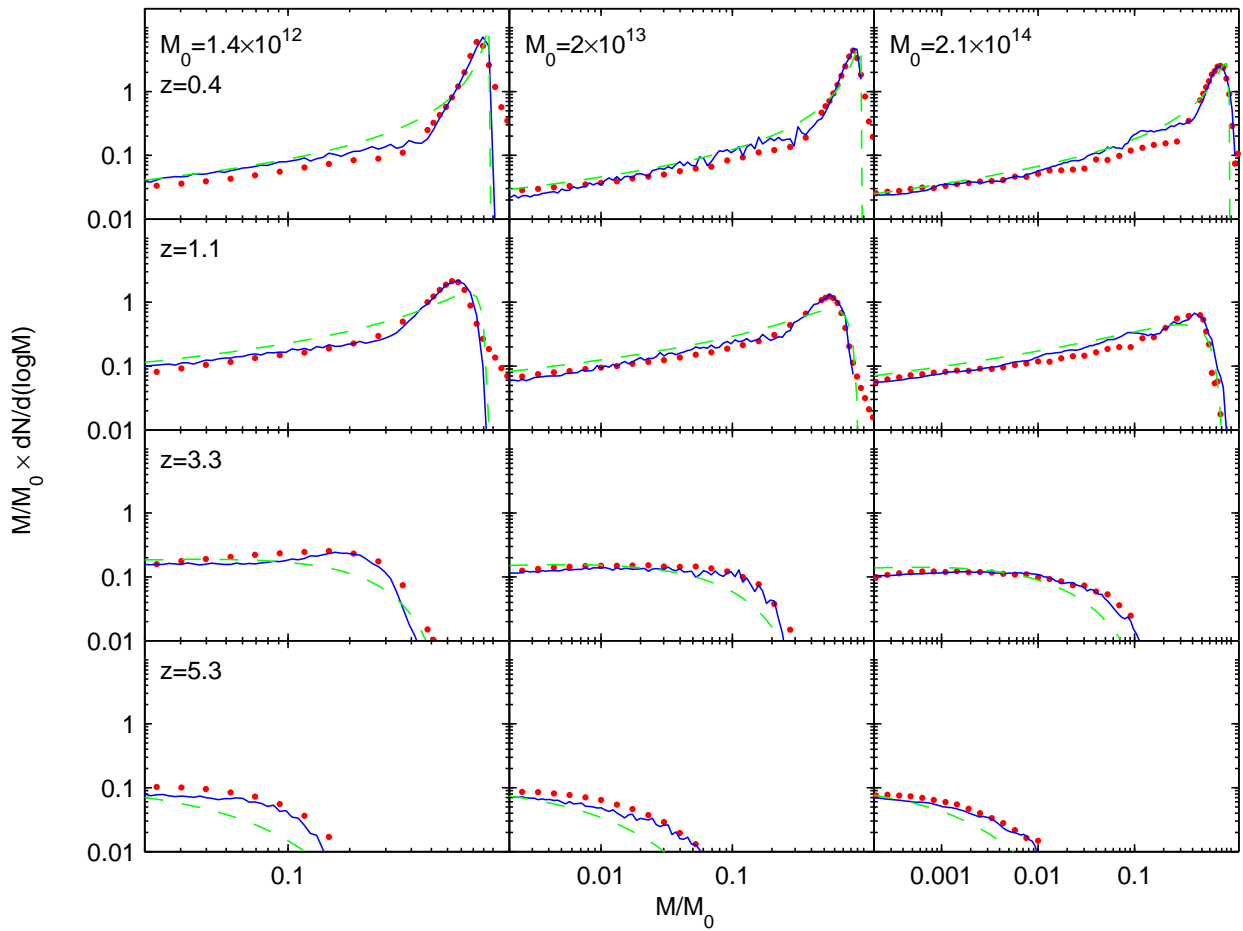,width=220mm,bbllx=1mm,bblly=85mm,bburx=188
mm,bbury=195mm,clip=}}
\caption{The progenitor number density $\dd N/\dd M$ at redshift $z$
for haloes of mass $M_0$ at $z_0=0$.  The data from the Millennium Run (filled circles) are compared to the results from merger trees generated by our algorithm (solid curve). We used 20,000, 3,000 and 600 random trees in the three mass bins, respectively. Also shown is the EPS prediction based on \equ{condprobM} (dashed curve). }
\label{fig:dNdM}
\end{figure*}

The full merger tree involves much more than the MP history. The
progenitors in each time step may involve one or more secondary
progenitors above the minimum mass, and the mutual probabilities could in principle be rather complicated.  Here, we find based on a simple
symmetry rule that to a good accuracy all the progenitors can be drawn from the same kernel distribution function $K_a$, a generalization of
the $K_1$ used for the MP.

Recall that the MP kernel distribution $K_1$ predicts the value of
$\dS_1$. The mass of the MP should then be computed by $M_{1} =M(S_0+\dS_1)$. The mass available for the additional progenitors is $M_0-M_1$. The simplest approach might be to use $K_1$ again, this time in reference to $M_0-M_1$, in order to obtain $\dS_2$. The mass of the second progenitor can then be computed by $M_{2}=M(S_a+\dS_2)$, where $S_a=S(M_0-M_1)$. This process can be repeated in order to draw all the progenitors in each time-step. This automatically guarantees that the mass of all progenitors will be smaller than $M_0$. Surprisingly, this straightforward algorithm gives good results for haloes of mass $\lesssim 10^{12}\,\hmsun$. It turns out that the accuracy of the results is improved with a little modification, defining $S_a = S(0.95M_0-M_1)$ for the second progenitor,
$S_a = S(0.95M_0-M_1-M_2)$ for the third progenitor,
and so on. The results of this simple algorithm practically coincide with the results of the algorithm described below at $M_0=10^{12}\, \hmsun$.

Encouraged by the success of the simple algorithm for small mass
haloes, we wish to generalize it to more massive haloes, where the
number of progenitors per time-step may be larger and the progenitors
may be small. When generating the $n$'th progenitor in a given
time-step, define
\begin{eqnarray}
\label{eq:mleft}
\mleft \!\!\!\!&=&\!\!\!\! fM_0 - \sum_1^{n-1} M_i\;, \\ \nonumber
\sleft \!\!\!\!&=&\!\!\!\! S(\mleft) \;,
\end{eqnarray}
where $M_i$ is the mass of the $i$'th progenitor and $f=0.967 - 0.0245\log_{10}(S_0)$, except for the first progenitor where $f=1$. The original kernel $K_1$ of \equ{K(dS)}, with the moments $\mu_k$ and $\sigma_k$, is replaced by the log-normal function $K_a(\dS|S_0,\sleft)$, with the moments
\begin{eqnarray}
\label{eq:K_a}
\mu_a \!\!\!\!&=&\!\!\!\! \mu_k + (\sleft-S_0)(2.70 -4.76s +2.9s^2) \;,\\
\nonumber
\sigma_a \!\!\!\!&=&\!\!\!\! \sigma_k + (\sleft-S_0)(0.104+0.118s) \;,
\end{eqnarray}
where $s \equiv \log_{10}(S_0)$.

The algorithm for constructing a full merger tree, above a minimum halo mass $M_{\rm min}$, is thus as follows:
\begin{enumerate}
    \item \label{item:a} Draw a random $\dS_1$ from the log-normal
    distributed $K_a$ defined in \equ{K_a}
    (note that $\sleft=S_0$ gives $K_1$ from \equnp{K(dS)}).
    \item Compute the MP mass $M_{1}=M(S_0+\dS_1)$.
    \item \label{item:c} Compute $\mleft$ and $\sleft$ using
    \equ{mleft}. If $\mleft$ turns out larger than $M_1$, have $\mleft=M_1$. In this way, $M_1$ is guaranteed to be the most massive progenitor.
    \item \label{item:d} Draw a random $\dS_2$ using the same
    $K_a$ of \equ{K_a} and compute $M_{2}=M(S_{\rm left}+\dS_2)$.
    \item \label{item:e} If $M_2<M_{\rm min}$, re-generate $M_2$ by repeating
      step \ref{item:d}.
    \item Repeat steps \ref{item:c}-\ref{item:e} until $\mleft$ is smaller than
    $M_{\rm min}$.
\end{enumerate}
%

The above procedure is very efficient. The code we used for constructing the trees is able to produce full trees from $z=0$ up to $z\sim8$ at a rate of $\sim 10^{5} \times M_{\rm min}/M_0$ per second. For example, with $M_0=10^{14}\hmsun$ and $M_{\rm min}=1.72 \times10^{10}\hmsun$ as in the MR, a typical tree is constructed at 0.02 seconds using our $\sim1$ GHz computer. This tree has a total number of 15,000 progenitors on average.

\Fig{dNdM} displays the progenitor mass function $\dd N/\dd M$. The
quantity plotted is actually $\dd N/\dd \log M$ times $M/M_0$, so that each equal log interval along the x-axis contributes to the total mass $M_0$ in proportion to the corresponding value on the y axis. The
results from the MR are compared to the results from merger trees that were generated using our algorithm. When the time step is not
sufficiently large ($z=0.4$), our algorithm shows some deviations from the simulation results because of the non-Markov effects in the
latter are still non-negligible. At higher $z$ the fit is better, but
not perfect. Deviations as high as $\sim 20\%$ can be seen at low $z$
for massive haloes and at high $z$ for small haloes. This is
significantly better than the EPS predictions also shown in \fig{dNdM}, which show deviations of a factor $\sim 2-3$ in many
cases. It is likely that even better results can be obtained after a
more elaborate tuning of our model parameters, though the accuracy is
limited by the imperfections in the time invariance even when $\dW$ is used, and the limitations of a Markov model discussed in \se{markov_limitation}.

The results of \fig{dNdM} can be compared to the results reported in parallel by \citet{Cole07}, who provide global fitting function for $\dd N/\dd M$ from different FOF merger trees of the MR.

\begin{figure}
 \centerline{ \hbox{ \epsfig{file=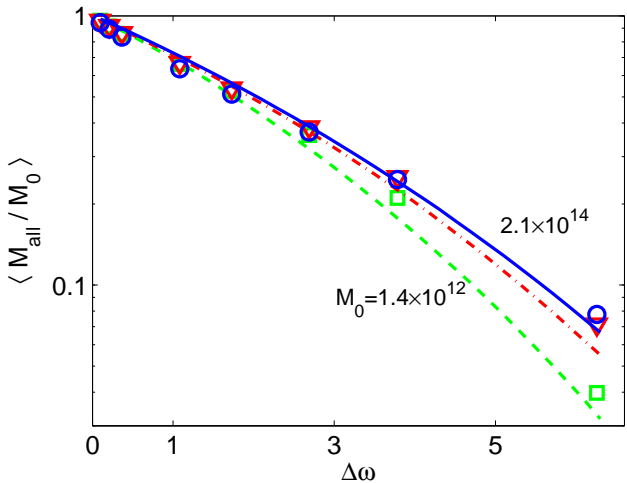,width=9cm} } }
\caption{The average mass of all progenitors, $\mall$, as a function
of time step, $\dW$, for haloes of different masses as listed in Table \ref{tab:halo_masses}. The data from the simulation are marked by
squares, triangles and circles for $M_0=1.4\times10^{12}$,
$2\times10^{13}$ and $2.1\times10^{14}$ respectively. The
corresponding model predictions are marked by green dashed, red
dot-dashed and blue solid lines respectively.}
  \label{fig:prog_sum}
\end{figure}

\Fig{prog_sum} shows the average mass encompassed in {\it all\,} the
progenitors above $M_{\rm min}$ as a function of $\dW$. This is the
average sum $\mall\equiv\sum M_i$, or the integral of the mass
function of \fig{dNdM} --- a quantity of interest for several
applications \citep{Navarro97,Neistein06,Neto07}. The results from the MR are compared to the averages from many realization of merger trees
generated by our algorithm. An interesting feature of $\mall$ is that
for small $\dW$ it shows a rather weak dependence on halo mass. This
implies that during that epoch all haloes gain the same fraction of
their mass via smooth accretion below $M_{\rm min}$, despite the fact
that $M_0/M_{\rm min}$ varies. This is related to the fact that the
progenitor mass function $\dd N/\dd M$ has a similar tail at low
masses for all halo masses. The last point can be seen in
fig.~\ref{fig:dNdM}, as histograms of different $M_0$ but with the
same $z$ are all similar at the low mass end.

The algorithm presented above has been empirically tuned to reproduce the distribution of the MP mass $P_1$ and the total mass function
$\dd N/\dd M$, at big enough time-steps ($\dW\gtrsim0.8$). Lacking an obvious physical motivation, it is not guaranteed a priori to also recover the correct full joint distribution of progenitors. Nevertheless, we find that the algorithm manages to reproduce the second progenitor with adequate accuracy over a large range of halo masses. This is demonstrated in the next section, where the second-progenitor distribution is recovered quite accurately for $10^{13}\,\hmsun$ haloes. The algorithm may be less accurate for very small progenitors, $M\lesssim0.01M_0$, but these progenitors encompass only a small fraction of the mass at $\dW=0.1$, a few percents for $M_0=10^{13} \,\hmsun$.

The algorithm presented here can be compared to the one by \citet{Sheth99}, motivated by Poisson initial conditions. These
authors have developed an algorithm that is based on the notion that
mutually disconnected volumes inside a halo are mutually independent.
As a result, all the progenitors are drawn from the same probability
distribution depending on the density in each region. In our algorithm it is the mass steps $\dS_i$ that are almost independent, although the progenitor masses depend on each other through $\mleft$. Our current
study focuses on providing a recipe that reproduces the $N$-body
simulation data, but the symmetry that lies at the basis of our
successful algorithm may provide interesting clues that may lead to a
more physical model.


\section{Merger Rates}
\label{sec:merger_rates}

It is often very useful to extract from merger trees the merger rates
of haloes of different masses. This is a key ingredient in galaxy-formation models, where major merger are assumed to be an important channel for the formation of star bursts, spheroidal stellar systems and AGNs. The progenitor mass function $\dd N/\dd M$ addressed above is clearly not enough to constrain the merger rates \citep[e.g.][]{Sheth99}.

In the simulation, and in our Markov model, there are cases were a
halo has many progenitors per time-step, especially when the halo is
massive or when the time step is large. Because the order by which
progenitors merge may change the results, a complete self-consistent
treatment of merger rates should properly address all possible merger
sequences within a time-step. Here we limit our analysis to the joint
probability of the two most massive progenitors, $P_{1,2}( M_1,M_2 |M_0,\dW )$, with $M_1 \geq M_2$. In fact, we define here the merger-rate kernel to be similar to $P_{1,2}$, but with the additional simplifying constraint that no other mergers occur during the time-step $\dW$.

This approximation may admittedly be somewhat crude. On one hand,
we learn from the MR simulation that for $10^{13}\,\hmsun$ haloes and
$\dW=0.1$ about $\sim 90\%$ of the merger events with $M_i/ M_1\gtrsim 0.05$ involve only $M_1$ and $M_2$. On the other hand, the residual mass in all other progenitors is on average about one third of $M_2$, i.e., not negligible. Had we merged these small progenitors with $M_2$ prior to its merger with $M_1$, the change in $M_2$ would have induced a non-negligible change in the quoted merger rate for $M_1$ and $M_2$. Our approximation for this merger rate becomes better if the smaller progenitors merge first with the much larger $M_1$, or merge after the $M_1$-$M_2$ merger altogether.

\begin{figure}
 \centerline{ \hbox{ \epsfig{file=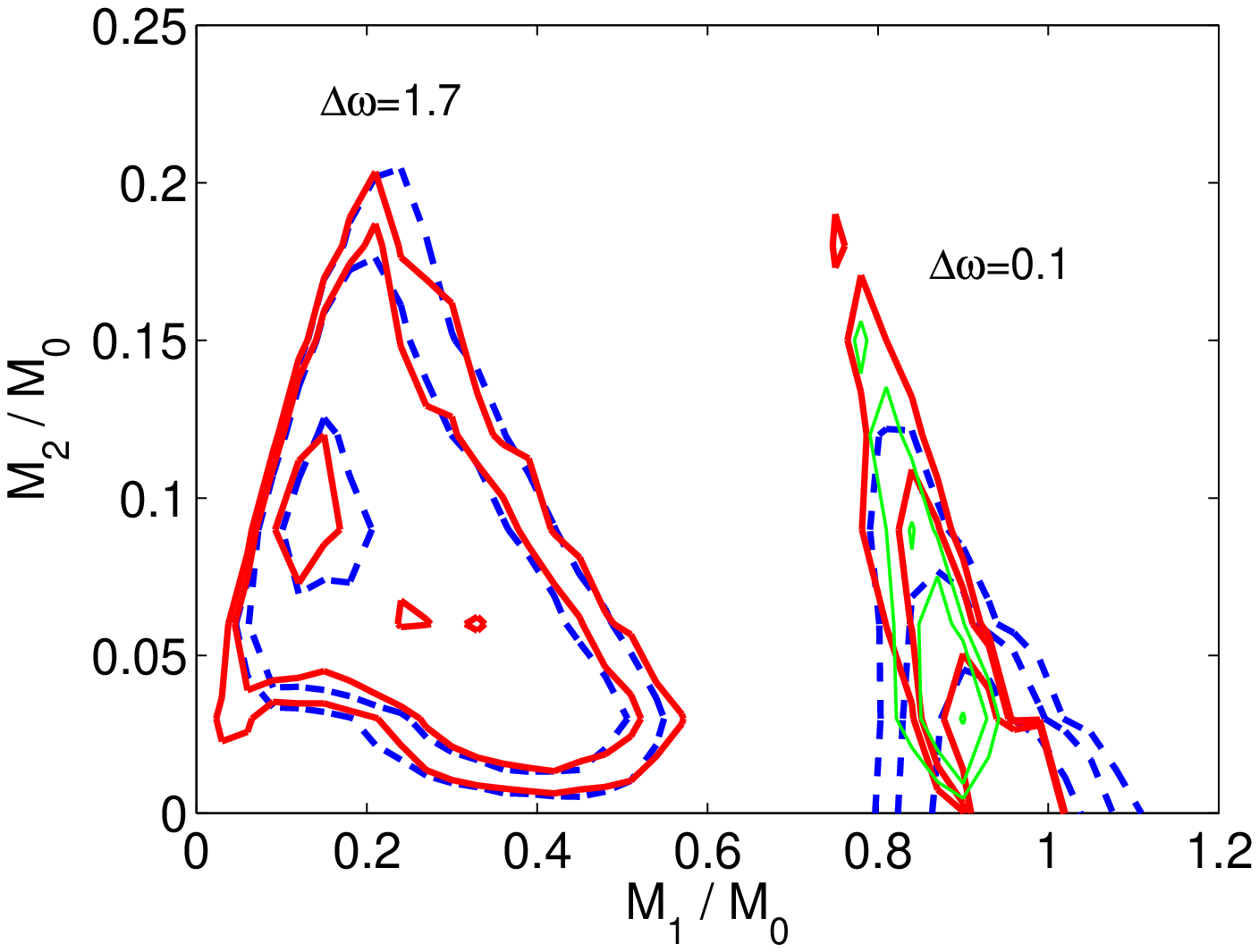,width=9cm} } }
 \centerline{ \hbox{ \epsfig{file=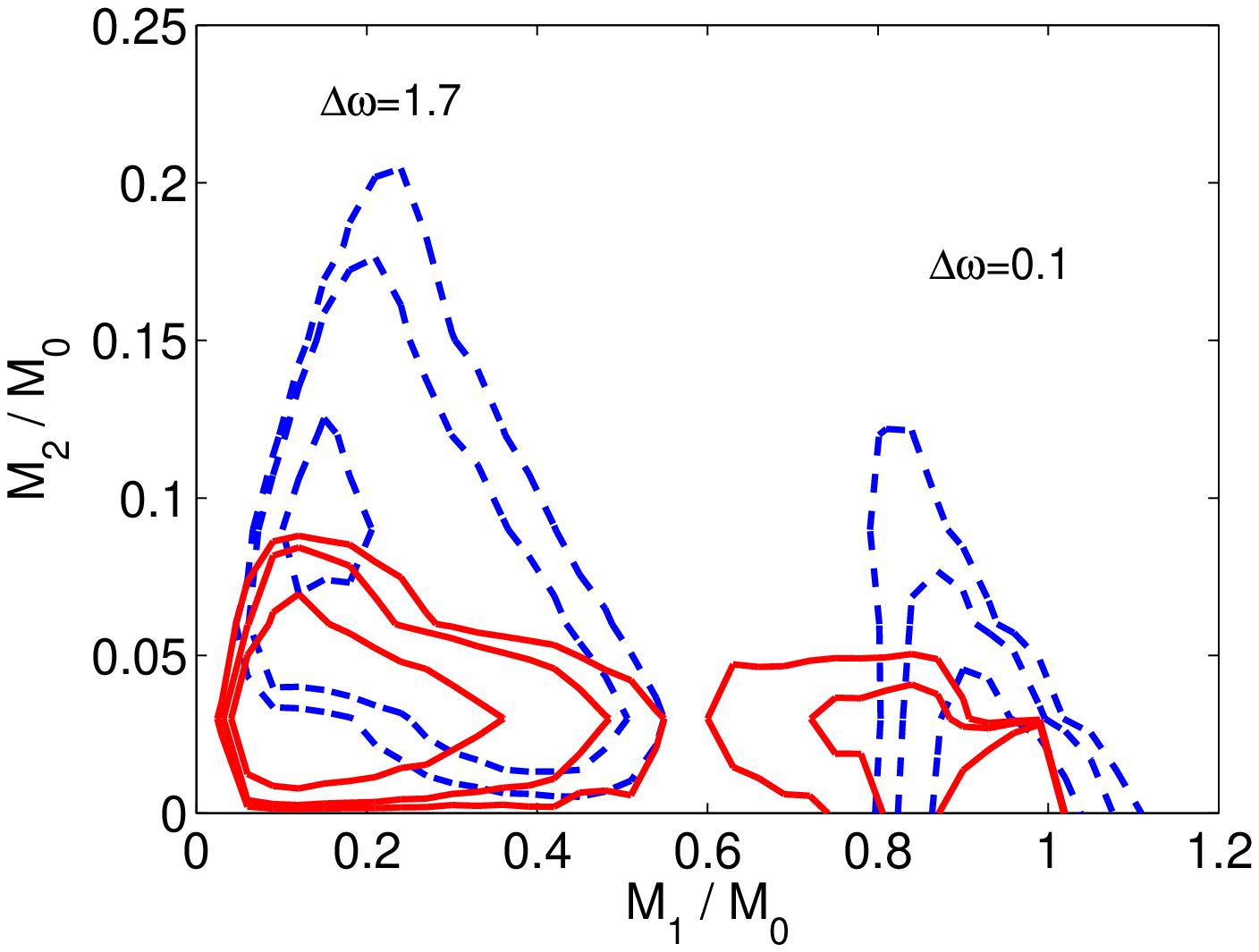,width=9cm} } }
\caption{Joint distribution of the two most massive progenitors
$P_{1,2}$. Each panel shows two snapshots in time, at $\dW=0.1$ and
1.7. The contour levels are at $P_{1,2}=5,10,30 \;M_0^{-2}$. The
results from the MR for the mass bin of $2\times10^{13}\,\hmsun$ are
shown as dashed blue contours. \emph{Upper panel:} The thick solid red contours refer to realizations of merger trees generated by our
algorithm. The thin green contours are the analytic approximation of
\equ{merger_rate}. \emph{Lower panel:} The solid red contours refer to realizations of EPS merger trees using the algorithm of
\citet{Somerville99a}.}
\label{fig:p12}
\end{figure}

\Fig{p12} compares $P_{1,2}$ from $10^4$ merger trees generated by our algorithm with MR trees for a halo of mass $2\times10^{13}\,\hmsun$
and for $\dW=0.1$ and $1.7$. Our algorithm nicely fits the simulation
at big $\dW$. The fit is only qualitative at the small time step,
$\dW=0.1$. We know already that deviations along the $M_1$ axis are
expected due to the non-Markov behaviour of the MP (\se{markov_model}), and we will see below (\se{markov_limitation})
that the deviations along the $M_2$ axis are also unavoidable.

The lower panel of \fig{p12} shows results from EPS merger trees
constructed using the standard algorithm of \citet{Somerville99a}. We
see that this algorithm underestimates the mass of the second
progenitor at all time steps. This discrepancy was not obvious in \citet[][Fig.~9]{Somerville00}, probably because of the rather small ratio of $M_0/M_{\rm min} \sim 40$ used there (in comparison with $\sim 1000$ here). Different algorithms based on EPS may yield different results, and our preliminary study indicates that it would be possible to develop an EPS algorithm in a spirit similar to our current model such that its $P_{1,2}$ will provide a better fit to the $N$-body results.

Our results could be compared to the estimate by \citet{Lacey93} for mergers in the limit of infinitesimal time steps within the framework of EPS. They assumed that in this limit merger events are binary, so $M_2$ is fully determined by $M_1$ and $M_0$. The assumption of binary mergers in small time-steps can be tested in \fig{p12}, where the distribution of $P_{1,2}$ for $\dW=0.1$ is clearly peaked near the line $M_1+M_2=M_0$. This may explain why \citet{Lacey94} found a good match between their EPS formula and results from $N$-body simulations. However, this approach is not fully consistent. Using infinitesimal time steps $P_{1,2}$ actually converges to a Dirac delta function about $(M_1,M_2)=(M_0,0)$, and therefore cannot be used to predict $P_{1,2}$ at finite time steps. Moreover, we show in Appendix \ref{sec:EPS_mergers} that in the limit of small time steps the EPS formalism does not converge to binary mergers. This may explain why the formula of \citet{Lacey93} fails to yield the correct symmetry between the two merging progenitors
\citep{Benson05}.

Our algorithm can be expressed in terms of an analytic estimate for
$P_{1,2}$. We assume that the second progenitor drawn is also the
second most massive. In this case:
\begin{eqnarray}
\label{eq:merger_rate}
\lefteqn{ P_{1,2}(M_1,M_2|M_0,\dW_0=0.1) = } \\
\nonumber & & K_a(\dS_1|S_0,S_0) \cdot K_a(\dS_2|S_0,\sleft) \frac{\dd
S(M_1)}{\dd M} \frac{\dd S(M_2)}{\dd M}  \;,
\end{eqnarray}
where $\dS_1 = S(M_1)-S(M_0)$, $\dS_2 = S(M_2)-\sleft$, $\sleft$ is
defined by \equ{mleft} with $n=2$, and $K_a$ is given by \equ{K_a}.
This approximation is shown for the short time step in the upper panel of \fig{p12}. We see that the analytic expression provides a crude approximation for the results from the merger trees constructed by the full algorithm and the MR simulation for $M_2/M_0\gtrsim0.05$. The analytic approximation apparently fails at lower values of $M_2$. This is because the second progenitor drawn is no longer necessarily the second most massive.

Our approximate formula for the merger rate $P_{1,2}$, \equ{merger_rate}, is time-invariant; it holds for any measurement
redshift $z_0$ where $M_0$ is identified. Its change with time becomes apparent only when the rate is expressed with respect to a unit of
time rather than $\omega$, i.e., the merger rate is $\propto
\dot\omega$. In the $\Lambda$CDM cosmology used here $\dot\omega
\propto (1+z)^m$ where $m$ varies from $\sim 2.2$ at low redshift to
an asymptotic value of 2.5 at high redshift (see \equnp{dwdt}). Early
studies of merger rates in $N$-body simulations found somewhat higher
values in the range $2.5 \lesssim m \lesssim 3.5$ \citep[e.g.][]{Governato99,Gottlober01}. It is not clear at this point how accurate these $N$-body estimates are. If future measurements of
$N$-body merger rates indeed turn out different from our time-invariant predictions, one can think of several potential reasons for such deviations. First is the non-Markov nature of $N$-body
merger rates at small time steps. If, for instance, it is due to the
finite relaxation time after a merger, and if this time is associated
with the halo dynamical time that varies with redshift, then the
non-Markov effects may vary with redshift. Second is the imperfection of the time invariance when using $\dW$. Thirdly, the deviation may arise from the differences between $P_{1,2}$ and the actual merger rate, where multiple mergers are not negligible.


\section{Markov and non-Markov Phases}
\label{sec:markov_limitation}

Despite the fact that the $N$-body trees are not Markovian at small
time-steps, we saw that our Markov algorithm manages to reproduce
many of the tree properties across big time-steps, including the MP
distribution, the progenitor mass function, the merger rates and the
total mass in all the progenitors. We also saw some inaccuracies of
the Markov model in reproducing the tree properties and merger
rates. In particular, \fig{p12} indicates that while the average mass
of the main progenitor is reproduced quite accurately, the average mass of the second progenitor is inaccurate. Here we address additional non-Markov aspects of the small-progenitor behaviour.

\begin{figure}
 \centerline{ \hbox{ \epsfig{file=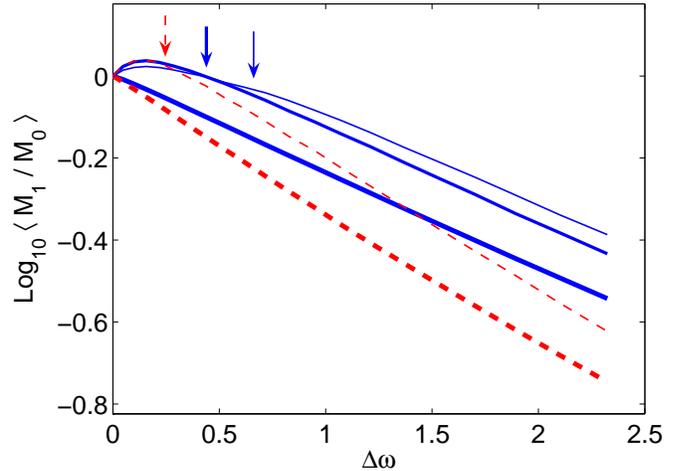,width=9cm}}}
\caption{Average MP history for haloes of a given mass identified at
$z_0=0.4$, grouped according to their future between $z=0.4$ and
$z=0$. The three solid blue curves refer to haloes of mass
$1.4\times10^{12}\;\hmsun$. Shown is the evolution of all the haloes
(thick line), those that end up as $2\times10^{13}$ haloes at $z=0$
(medium line), and those that end up as $2.1\times10^{14}\,\hmsun$ at
$z=0$ (thin line). The two dashed red curves are for haloes of
$1.5\times10^{13}\;\hmsun$ at $z_0=0.4$, where the thick line is all
haloes, and the thin line is the average of haloes that end up as
$2.1\times10^{14}\,\hmsun$ haloes at $z=0$. All progenitors show the
familiar growth at a uniform rate at early times, while those that end up as massive haloes at $z=0$ show mass loss during the last $\dW \sim 0.1$ before $z_0=0.4$. The point were the slope of each curve starts
deviating from the slope at high redshift is marked by an arrow.}
  \label{fig:non_markov_progs}
\end{figure}

In \fig{non_markov_progs} we show average MP histories of haloes of a
given mass as identified at $z_0=0.4$, grouped according to their {\it future\,} evolution from $z=0.4$ to $z=0$. We see that on average
those haloes that will end up as part of a more massive halo at $z=0$
have already suffered an abnormally slow growth starting $\dW \sim
0.2$ or more before $z_0=0.4$. This actually turns (on average!) into
a period of mass loss during the last $\dW\sim0.1$ just before $z_0$.
This is clearly a non-Markov behaviour. It is probably due to
tide-limited accretion at the vicinity of massive haloes and tidal
stripping once passing inside such haloes \citep{Diemand07, Desjacques07,Hahn07}. At early times, e.g., more than $\dW\sim0.5$ prior to $z_0$, the growth rate of these special haloes is similar to the average of all the haloes, but the value of $M_1$ at any given time is $\sim1.5$ times higher. The transition from average growth rate to a suppressed growth rate can be identified, as marked by the arrows in \fig{non_markov_progs}. We could interpret this as transition from a Markov to non-Markov behaviour.

The non-Markov effects limit the accuracy of our model.
In particular, the progenitor mass function of \fig{dNdM} do not approach a Markov behaviour even at high redshift, in the sense that the success of the Markov model in one time step does not guarantee its success in other time steps. This is because many of the progenitors present at a given redshift are about to merge into a much bigger halo a short time later, and are therefore subject to mass loss that induces a non-Markov behaviour. This explains why our model fits for $\dd N/\dd M$ show non-negligible deviations from the simulated mass functions.

The fact that the mass of some haloes is not monotonically increasing
with time is doomed to complicate the interpretation of the merger
rates, even if the mergers are counted properly in a given time-step
in the simulation. It is not obvious how to formulate a self-consistent and time-invariant recipe for merger rates given that
these haloes were actually more massive at some point in the past.
Nevertheless, despite the shortcomings of any Markov model, it
provides a sensible basis for a self-consistent definition of merger rates. For one thing, the merger rates of a non-Markov model are likely to have an undesired dependence on the length of the time step chosen for the tree. Our Markov model tends to overestimate the mass of the secondary progenitors at small time-steps (seen in \fig{p12} as a stretching of the model contours toward higher values of $M_2$), thus approximately compensating for the opposite effect due to mass loss. This is an outcome of the model tuning, designed to fit the data of $\dd N/\dd M$ at large time steps.

The non-Markov effect seen in \fig{non_markov_progs} is related to
the environment dependence of the assembly time for distinct haloes
\citep{Gao05,Harker06} through the natural correlation between the
future halo mass and its current environment. \Fig{non_markov_progs}
demonstrates why the formation redshift of a halo that resides in a
high-density environment is higher than average. With higher mass loss prior to $z_0$, the ``formation time", when the MP was $M_1=0.5\,M_0$, is clearly earlier. This is quantified in \citet{Hahn07} and \citet{Desjacques07}. We see that the environment effect is at least partly associated with haloes in their non-Markov phase where they are about to merge into bigger haloes.  We may therefore expect a weaker or no environment effect for haloes in their Markov phase of typical monotonic growth, before the transition marked by an arrow in \fig{non_markov_progs}. This division into haloes in the Markov phase versus those in the non-Markov phase might be a natural way to divide the halo population, more physical than the standard division to
``distinct haloes" versus ``subhaloes" based on the virial radius.

Simulations and observational data indicate that, unlike dark haloes, the stellar galaxies that reside in them tend not to show a significant environment dependence \citep{Croton07,Tinker07}. While, as seen in \fig{non_markov_progs}, the mass-loss induced environment effect occurs mainly at late times, the stellar systems might have crystalized as compact systems at earlier times, which makes them less subject to tidal effects.


\section{Discussion}
\label{sec:discuss}

We addressed the statistics of dark-matter merger trees, as extracted
from the Millennium $N$-body simulation. We demonstrated that the time and mass variables of the EPS formalism, $\omega(t)$ and $\sigma^2(M)$, are indeed the natural variables for describing merger trees in a time-invariant way, at an accuracy level of a few percent. It may be interesting to explore different ways to define haloes in an attempt to improve the time invariance of the statistics.
This includes, for example, different systematic time variations of the overdensity or linking length used to define the haloes.
It may also be worthwhile to test the time invariance in an idealized
Einstein-deSitter cosmology with a power-law power spectrum, where there could be a better chance to isolate the non-Markov contribution to any violation of self-similarity.

The log-normal nature of the distribution of the main progenitor
as a function of $\sigma^2(M)$ may be a clue to the physical origin of the statistics of merger trees. It may be associated with a product of multiple random processes through the central-limit theorem, but this is beyond the scope of our current analysis. It may be interesting to evaluate to what extent this log-normal behaviour is valid in different cosmological models, which could be interpreted as representing different density environments in a given $\Lambda$CDM cosmology. It should also be interesting to test the changes induced by using a different window function in the definition of $\sigma^2(M)$.

Despite the non-Markov nature of $N$-body trees, we showed that
they can be approximated by a Markov process of short time-steps that
reproduces the progenitor distribution at sufficiently long time-steps, $\dW > 0.5$. The average main-progenitor history is actually recovered accurately even at short time steps. In addition, the distribution of full $N$-body merger trees can be reproduced by a similar probability distribution function for all the progenitors.
The progenitors are drawn one after the other from the mass left in
the descendant halo after subtracting the progenitors chosen so far.
We demonstrated that the joint distribution of the two most massive
progenitors is reproduced quite accurately. This algorithm can thus be used to construct semi-analytic merger trees that resemble the statistics of $N$-body merger trees better than any previous algorithm. It is in principle applicable at any desired mass resolution and in any cosmological model. However, the non-Markov features of the merger trees limit the accuracy. Preliminary tests indicate that a similar model can possibly be developed for merger trees based on the EPS formalism.

Extracting merger rates from the simulation is a non-trivial task. First, with several progenitors in each time-step, the order by which they merge matters for the merger rates and should be properly modeled. Second, the non-Markov suppression of growth rate, e.g., due to tidal effects makes the progenitor mass just prior to a merger differ from the masses as extrapolated from the same progenitors at high redshift. Still, we deduce from our Markov algorithm a simple
approximation to the merger rate kernel for the two most massive
progenitors. Once applied over short time-steps, it reproduces the
high-$z$ progenitor mass with good accuracy. The time invariance of
our algorithm implies that the merger rates evolve in time in
proportion to $\dot\omega \sim (1+z)^m$, where $m$ ranges from $\simeq 2.2$ at low $z$ to 2.5 at high $z$. This time invariance may be invalidated by non-Markov effects that evolve with time, such as
the dynamical time of haloes. For all the reasons above, the success
of our approximate merger rates in reproducing the actual $N$-body
merger rates is yet to be evaluated.

Our algorithm suggests a natural distinction between Markov and
non-Markov haloes, or phases in the evolution of a halo. The
Markov phase is when the halo grows monotonically in time in a rate
close to the average rate, before it is suppressed, presumably by
tidal effects in the neighborhood of massive haloes. The population of Markov haloes should not show the environment dependence of halo
formation time \citep{Gao05,Harker06}. This distinction would rely on
non-local halo properties, such as its proximity to more massive
structures. A practical definition of non-Markov haloes may be
those that will become subhaloes of a bigger halo in the next time
interval corresponding to $\dW \sim 0.5$. However, this particular
tentative definition has the undesired effect of smoothing the time
resolution of the tree. Working out a similar distinction without
suppressing the tree resolution is an interesting challenge for future work.

\textsc{Matlab} and \textsc{c} codes of the algorithm presented in this paper are available on the web at http://www.phys.huji.ac.il/$\sim$eyal$\underline{\;\;}$n/merger$ \underline{\;\;}$tree/  and can be used as a black box for constructing merger trees.


\section*{Acknowledgments}

The Millennium Simulation databases used in this paper and the web
application providing online access to them were constructed as part
of the activities of the German Astrophysical Virtual Observatory. We
thank Gerard Lemson for his help in using these databases. We acknowledge stimulating discussions with Christian Maulbetsch, Vincent Desjacques, Noam Libeskind, Elad Zinger, Joanna Woo and Daniel Darg.
This research has been supported by ISF 213/02, by GIF
I-895-207.7/2005, by the Einstein Center at HU, and by NASA ATP
NAG5-8218.

\bibliographystyle{mn2e}
\bibliography{eyal_1}

\appendix

\section{Computing $\omega$ and $S$}
\label{sec:app_w_s}

In this section we describe in detail how we compute the natural
variables, $\omega(z)$ and $S(M)$. The cosmological parameters in the
MR are $(\Omega_{\Lambda}, \Omega_{\rm m}, \sigma_8, h) =(0.75, 0.25,
0.9, 0.73)$. We use the standard power spectrum $P(k)=kT^2(k)$, with
the transfer function \citep{Bardeen86}
\begin{eqnarray}
\lefteqn{T(k) = \frac{\ln(1+2.34q)}{2.34q}\times } \\
\nonumber & & \left[1 + 3.89q + (16.1q)^2 + (5.46q)^3 + (6.71q)^4
\right] ^{-1/4} \,.
\end{eqnarray}
Here $q = k/\Gamma$, with $k$ in $h$Mpc$^{-1}$, and $\Gamma=0.169$ is
the power spectrum shape parameter chosen to best fit the CMBFAST
model \citep{Seljak96} used in the MR.

We use the definition of $S(M)$ from \citet{Lacey93} as the variance
of the density field smoothed with a spherical top-hat window function of a radius that on average encompasses a mass $M$ in real space. In
practice we use the fitting function given by \citet{vdBosch02b}:
\begin{equation}
S(M) = u^2 \biggl[ \frac{c_0\Gamma}{\Omega_m^{1/3}}M^{1/3} \biggr] \cdot
\frac{ \sigma_8^2 } {u^2(32\Gamma)} \,,
\end{equation}
where $c_0=3.804\times 10^{-4}$, and $u(x)$ is an analytical function:
\begin{eqnarray}
\lefteqn{ u(x) = 64.087 \bigl[ 1 + 1.074x^{0.3} } \\
\nonumber & &  - 1.581x^{0.4} + 0.954x^{0.5} - 0.185x^{0.6} \bigr] ^
{-10} \,.
\end{eqnarray}

In order to compute $\omega(z)$ we use the recipe from
\citet{Navarro97}, which uses for the $\Lambda$CDM cosmology:
\begin{equation}
\omega = 1.6865\frac{ \Omega_z^{0.0055}}{D(z)} \;,
\end{equation}
where
\begin{equation}
\Omega_z =  \frac{ \Omega_m (1+z)^3}{\Omega_m
(1+z)^3+(1-\Omega_m-\Omega_\Lambda)(1+z)^2 + \Omega_\Lambda} \;.
\end{equation}
The linear growth rate $D(z)$ is computed by performing the integral:
\begin{equation}
D(z) = D_0 H(z) \int_z^\infty \frac{1+z_1}{H^3(z_1)}\dd z_1 \;,
\end{equation}
where $D_0$ is a constant set by the normalization $D(0)=1$. We
provide a practical approximation for $\omega(z)$,
\begin{equation}
\omega(z) = 1.260 \left[ 1+z +0.09(1+z)^{-1} + 0.24 e^{-1.16z}
\right]\,,
\end{equation}
which is accurate to better than 0.5\% at all redshifts for the
$\Lambda$CDM cosmology used here. As mentioned in section
\ref{sec:average_mmain}, the time derivative of $\omega$ can be well
approximated by:
\begin{eqnarray}
\dot{\omega} = -0.0470\left[ 1+z+0.1(1+z)^{-1.25} \right]^{2.5} \;
h_{73}\, {\rm Gyr}^{-1}\;,
\end{eqnarray}
where $h_{73}$ is the Hubble constant measured in units of 73
$\kms\,\Mpc^{-1}$. This is also good to better than 0.5\% at all
redshifts.


\section{Goodness of fit for the Main Progenitor}
\label{sec:app_mmain_fit}

\begin{figure}
 \centerline{ \hbox{ \epsfig{file=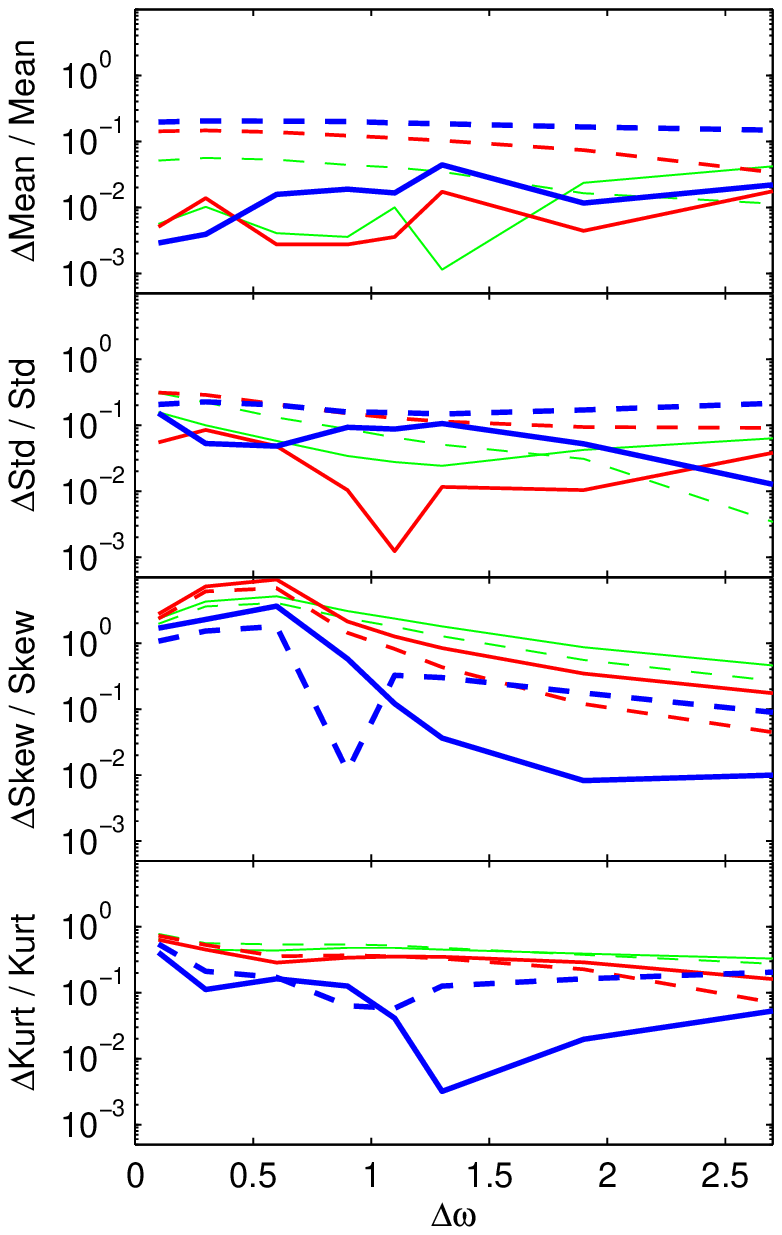,width=10cm} } }
\caption{Goodness of fit for the mass distribution of the main
progenitor $P_1(\dS_1|S_0,\dW)$. The two models of \se{main_prog}
are compared to the MR simulation. For each $M_0$ and $\dW$ we show the fractional deviation in the first four moments of the distribution. These moments are the mean, standard deviation, skewness, and kurtosis. The solid curves refer to the merger tree realizations using $K_1$ versus the simulation. The dashed curves refer to the global log-normal fit of \equs{logn_fit1}-(\ref{eq:logn_fit3}) in comparison with the simulation. The halo masses are in the three massive bins defined in table \ref{tab:halo_masses}, with the thickness of the line increasing with halo mass.}
  \label{fig:main_prog_fitq}
\end{figure}

In this Appendix we evaluate the quality of fit of the two models
presented in \se{main_prog} to the distribution of main-progenitor
mass in the MR simulation. First the straight-forward global fit for $P_1(\dS_1|S_0,\dW)$, \equs{logn_fit1}-(\ref{eq:logn_fit3}) is examined. The quality of this fit to the MR data is evaluated in \fig{main_prog_fitq} via the fractional deviations in the first four moments of $P_1$, the mean, standard deviation, skewness and kurtosis. The fit is reasonably good starting at $\dW\sim 0.8$, with deviations of $\sim 20\%$ in all moments, and with the skewness showing somewhat larger deviations. One reason for these deviations is the global nature of the fit, being performed once for all masses and times. Naturally, separate fits in limited mass ranges or epochs will yield better results. Another source of scatter is the limited sampling, which contributes an error of $\sim 0.5\%$ in the most massive bin.
For this bin, the sampling error is comparable to the deviations of the model average from the data.

Also shown in \Fig{main_prog_fitq} is the difference between the MP distribution as derived from $10^5$ merger-tree realizations of our Markov model with the kernel $K_1$ and the distribution in the MR simulation. Results of the same Markov model are also displayed in
\fig{main_prog_with_t}, where a specific halo mass is followed in time. The sources of scatter discussed above are also valid here. Additional scatter arises from the differences between the time-steps of our model and those in the simulation. As our model uses a fixed kernel with $\dW_0=0.1$, we generate predictions only at times which are integer multiples of $\dW_0=0.1$. We pick the closest possible output times from the MR, but this is only good to 10\% in $\dW/\omega$ at low $\dW$, and 0.5\% at high $\dW$. This source of scatter can be weakened by interpolation between time steps.


\section{Binary Mergers in EPS?}
\label{sec:EPS_mergers}

We define a ``binary merger" event by having exactly $M_1+M_2=M_0$ in
a given time step. We show here that this is not a valid limit in the
EPS formalism when the time step is infinitesimal. The number density
of progenitors as predicted by EPS is \citep[e.g.,][]{Lacey93}
\begin{eqnarray}
\label{eq:condprobM} \lefteqn{{{\dd}N \over {\dd}M}(M,z \vert M_0,z_0)
\; {\dd}M =}
\nonumber \\
& & {M_0 \over M} \; \frac{1}{\sqrt{2 \pi}} \; {\dW \over \dS^{3/2}}
\; {\rm exp}\left[-{\dW^2 \over 2 \dS}\right]  \; \left\vert {{\dd}S
\over {\dd M}} \right\vert \; {\dd}M \,.
\end{eqnarray}
When the progenitors of all masses down to $M \rightarrow 0$ are considered, this implies that any halo has an infinite number of progenitors at any previous time, not permitting a binary event even at small time-steps. Only when a minimum halo mass $M_{\rm min}$ is imposed can a binary merger occur. However, we show below that the
mass in ``progenitors" below $M_{\rm min}$, which one may term
``smooth accretion'', $M_{\rm acc}$, is never negligible compared to $M_2$.

The average $M_{\rm acc}$ is obtained by integrating $\dd N/\dd M
\times M$ between 0 and $M_{\rm min}$,
\begin{equation}
\frac{ \langle M_{\rm acc} \rangle }{M_0} = {\rm erf} \left[
\frac{\dW}{\sqrt{2S_{\rm min}-2S_0}} \right] \;,
\end{equation}
where $S_{\rm min}=S(M_{\rm min})$ and $S_0=S(M_0)$. It has been shown in \citet{Neistein06} that the main-progenitor distribution at small time-steps equals $\dd N/\dd M$ for $M>M_0/2$, with a small tail extending to low masses $M<M_0/2$. Consequently, the probability of the second progenitor, $P_2$, roughly equals $\dd N/\dd M$ in the range $M_{\rm min}<M<M_0/2$. Since the latter is always a slight overestimate, we use it for an upper limit to the average mass of the second progenitor. Integrating $\dd N/\dd M \times M$ we obtain
\begin{equation}
\frac{ \langle M_2 \rangle }{M_0}\leq{\rm erf} \left[
\frac{\dW}{\sqrt{2S_2-2S_0}} \right] - {\rm erf}\left[
\frac{\dW}{\sqrt{2S_{\rm min}-2S_0}} \right] \;,
\end{equation}
where $S_2=S(M_0/2)$.

In the limit of small time-steps, $\dW\rightarrow0$, using ${\rm erf}(x)\rightarrow 2x/\sqrt{\pi}$ as $x\rightarrow0$, we get
\begin{equation}
\frac{ \langle M_2 \rangle }{ \langle M_{\rm acc} \rangle } \leq
\sqrt{\frac{S_{\rm min}-S_0}{S_2-S_0}} -1 \,.
\label{eq:M2_Macc}
\end{equation}
For the $\Lambda$CDM cosmology used here, and with the minimum mass of $1.72\times10^{10}\,\hmsun$ in the Millennium simulation, this upper limit varies between 2 and 6.5 for haloes of mass $10^{12}$ to $10^{14}\,\hmsun$. The actual value of $\langle M_2 \rangle / \langle M_{\rm acc} \rangle$ is somewhat lower, and it may depend on the specific algorithm used to construct the trees.

One may argue that in \equ{M2_Macc} we can take $S_{\rm min}$ to
infinity as $M_{\rm min}$ goes to zero, so $\langle M_2 \rangle /
\langle M_{\rm acc} \rangle$ will approach infinity as well.
Apparently, this procedure may seem to eliminate the minimum mass and make the accreting mass vanish such that the limit of binary mergers is reproduced. We want to emphasize that this limit is not well defined in EPS. It can be shown that in the limit $S \propto \dS \rightarrow \infty$ and $\dW \rightarrow0$ \equ{condprobM} approaches $\dW M^{-1} S^{-1.5} \dd S$. This expression actually depends on the way by which each variable approaches its limit, so the procedure can practically yield an arbitrary result.

Thus, the accretion mass is always comparable to $M_2$, even when they both vanish linearly with $\dW$. This implies that the merger rate as
computed by \citet{Lacey93} (their eq.2.17) is not self-consistent
within the EPS formalism and may therefore be invalid. It may explain
why \citet{Benson05} found this merger rate problematic. The situation is different in merger trees constructed from $N$-body simulations,
where every given halo has a finite number of particles, thus
introducing a natural $M_{\rm min}$ at the particle mass. In this case, binary mergers occur in the limit of small time-steps, as there is no smooth-accretion component.


\label{lastpage}

\end{document}